# Hybridization-Induced Gapped and Gapless States on the Surfaces of Magnetic Topological Insulators

Xiao-Ming Ma[1*], Zhongjia Chen[1*], Eike F. Schwier[2*], Yang Zhang[3*], Yu-Jie Hao[1], Rui'e Lu[1], Jifeng Shao[1], Yuanjun Jin[1], Meng Zeng[1], Xiang-Rui Liu[1], Zhanyang Hao[1], Ke Zhang[2], Wumiti Mansuer[2], Shiv Kumar[2], Chunyao Song[4], Yuan Wang[1], Boyan Zhao[1], Cai Liu[1], Ke Deng[1], Jiawei Mei[1], Kenya Shimada[2], Yue Zhao[1], Xingjiang Zhou[4], Bing Shen[3#], Wen Huang[1#], Chang Liu[1], Hu Xu[1], Chaoyu Chen[1#]

[1] Shenzhen Institute for Quantum Science and Engineering (SIQSE) and Department of Physics, Southern University of Science and Technology (SUSTech), Shenzhen 518055, China.

[2] Hiroshima Synchrotron Radiation Center, Hiroshima University, Higashi-Hiroshima, Hiroshima 739-0046, Japan

[3] School of Physics, Sun Yat-Sen University, Guangzhou 510275, China

[4] Institute of Physics and Beijing National Laboratory for Condensed Matter Physics, Chinese Academy of Sciences, Beijing 100190, China

[*]These authors contributed equally to this work.

[#]Correspondence should be addressed to B.S. (shenbingdy@mail.sysu.edu), W.H. (huangw3@sustech.edu.cn) and C.C. (chency@sustech.edu.cn)

**Abstract**

**The layered $MnBi_{2n}Te_{3n+1}$ family represents the first intrinsic antiferromagnetic topological insulator (AFM TI, protected by a combination symmetry $S$) ever discovered, providing an ideal platform to explore novel physics such as quantum anomalous Hall effect at elevated temperature and axion electrodynamics. Recent angle-resolved photoemission spectroscopy (ARPES) experiments on this family have revealed that all terminations exhibit (nearly) gapless topological surface states (TSSs) within the AFM state, violating the definition of the AFM TI, as the surfaces being studied should be $S$-breaking and opening a gap. Here we explain this curious paradox using a surface-bulk band hybridization picture. Combining ARPES and first-principles calculations, we prove that only an apparent gap is opened by hybridization between TSSs and bulk bands. The observed (nearly) gapless features are consistently reproduced by tight-binding simulations where TSSs are coupled to a Rashba-split bulk band. The Dirac-cone-like spectral features are actually of bulk origin, thus not sensitive to the $S$-breaking at the AFM surfaces. This picture explains the (nearly) gapless behaviour found in both $Bi_2Te_3$- and $MnBi_2Te_4$-terminated surfaces and is applicable to all terminations of $MnBi_{2n}Te_{3n+1}$ family. Our findings highlight the role of band hybridization, superior to magnetism in this case, in shaping the general surface band structure in magnetic topological materials for the first time.**

**Introduction:**

The discovery of topological insulators (TIs) in two dimensional (2D) and three dimensional (3D) material systems has triggered a paradigm evolution in condensed matter physics, in which the phase of matter and novel properties of quantum materials are explored based on the topology of electronic structure in reciprocal space [1-3]. Combination of the electronic topology and other degrees of freedom may lead to a variety of exotic phases of matter and physical responses such as axionic excitation [4-6], Majorana state and topological superconductivity [1,7]. More specifically, the interplay between magnetism and topology could generate various quantum states, such as antiferromagnetic TI (AFM TI) [8], the quantum anomalous Hall (QAH) state [9,10] hosting dissipationless chiral edge modes [11,12], the axion insulator displaying quantized magnetoelectric effects [4-6], and the magnetic Weyl and nodal-line semimetals [13,14] which harbour Fermi arc surface states and chiral anomalies [14].

The realization of magnetic TIs [15] demands crystalline materials with both magnetic order and electronic topology. Previously, the efforts were devoted to magnetically doped TIs and magnetic topological heterostructures [11,12,16–19], whose fabrication, measurement, and property optimization are quite challenging. Recently, $MnBi_2Te_4$ has arisen as the first AFM TI [20-26] with periodically ordered Mn atoms on well-defined crystallographic sites, hosting possible axion electrodynamics in condensed matter. The interlayer AFM with intralayer ferromagnetic (FM) order (c-axis A-type AFM) occurs below a Néel temperature of about $T_N = 24.6$ K [27-34]. In particular, given a high field to fully polarize the magnetic moments, quantized Hall effect arising from Chern insulator and the transition from axion to Chern insulator have been demonstrated in $MnBi_2Te_4$ in the 2D limit [35-38].

Considering its topology, AFM TI $MnBi_2Te_4$ can be classified by a topological $Z_2$ invariant [8], protected by a combination symmetry $S = \Theta T_{1/2}$, where $\Theta$ is time-reversal symmetry and $T_{1/2}$ is a lattice translation symmetry of the "primitive" lattice, both broken by the AFM order but their combination preserved. At the natural cleavage (001) surface of $MnBi_2Te_4$, $S$ is broken, rendering a sizable surface state gap, which is confirmed by early angle-resolved photoemission spectroscopy (ARPES) works [20,27,32,39]. This surface gap opening is crucial to realize the half-quantum Hall effect, which may aid experimental confirmation of $\theta = \pi$ quantized magnetoelectric coupling [8]. However, subsequent systematic $k_z$-dependent and high-resolution ARPES measurements clearly resolved a (nearly) gapless topological surface state (TSS) Dirac cone at the (001) surface of $MnBi_2Te_4$ [40-43]. This gapless feature remains intact in both paramagnetic (PM) and AFM phases, and is even robust against severe surface degradation [40,44], indicating additional topological protection from magnetic, structural or electronic complication. This may explain the necessity of high field for the observation of quantized Hall effect, even in thin films with odd number of layers [37,38].

In fact, $MnBi_2Te_4$ belongs to the ternary van der Waals compound series $(MnBi_2Te_4)_m(Bi_2Te_3)_n$

with [Te-Bi-Te-Mn-Te-Bi-Te] septuple (S) layers ([$MnBi_2Te_4$], SLs) and [Te-Bi-Te-Bi-Te] quintuple (Q) layers ([$Bi_2Te_3$], QLs) alternately stacking along the $c$ axis [45,46]. While $MnBi_2Te_4$ has been extensively studied recently [20-43], its sister compounds with higher $n$ number such as $MnBi_4Te_7$ ($n$ = 1) and $MnBi_6Te_{10}$ ($n$ = 2), remain less explored [47-58]. Structurally, the FM SLs are more separated in space as $n$ increases, reducing the interlayer AFM exchange coupling. Experimentally [48-55], $MnBi_4Te_7$ and $MnBi_6Te_{10}$ indeed show AFM transition at lower temperature (~13 and ~11 K, respectively) compared to $MnBi_2Te_4$. Much smaller magnetic field is needed to cause a spin flip transition. Importantly, an FM hysteresis was observed at low temperature, prerequisite for realizing the intrinsic QAH effect. However, similar to $MnBi_2Te_4$, controversial ARPES results exist regarding the TSSs gap in $n$ = 1 and 2 compounds. For S-termination, both gapped [50,58] and gapless [48,54,56,57] TSSs have been reported. Likewise, for Q-termination (SQ-and SQQ-terminations), both gapped [48-50,57] and gapless [54-56,58] TSSs were observed, too. As all types of terminations discussed here would be $S$-breaking in their AFM phases, the gapless TSSs violate the definition of AFM TI for all compounds. Thus, it is of critical importance to determine the underlying mechanism of gap opening for the intrinsic surface band structure in higher $n$ compounds.

In this work, combining ARPES, Density Functional Theory (DFT) and tight-binding (TB) analysis, we study $MnBi_6Te_{10}$ as an example to unveil the nature of the TSSs gap at different terminations. Using ARPES with focused Laser spot (~ 5 μm) and superb energy and momentum resolution (μ-Laser-ARPES [59]), four types of surface-driven band structures are spatially resolved, corresponding to the three terminations of $MnBi_6Te_{10}$, *i.e.*, S-termination, SQ-termination, SQQ-termination and one additional SQQQ-termination resulting from either vertical stacking disorder or degradation of the topmost SLs. This assignment is well supported by the DFT slab calculations based on different surface stacking configurations. The TSSs are gapped, but not found at the Dirac point, rather in the lower Dirac cone region for S-termination and in the upper Dirac cone region for the Q-terminations. From DFT slab calculations, this gap is found to open through the hybridization of the TSSs with a bulk band. We further show, via a TB model analysis taking into account the surface-bulk hybridization, one can indeed reproduce the experimental in-gap band. In particular, within the nonmagnetic phase, a new Dirac cone overtakes the original TI Dirac cone at the Γ point for all types of terminations. These Dirac-cone-like spectral features are actually of bulk origin, thus not sensitive to the $S$-breaking at the surfaces of AFM phase. Our findings provide a self-consistent picture to explain the (nearly) gapless behavior of the TSSs at all intrinsic terminations of $MnBi_4Te_7$ and $MnBi_6Te_{10}$ [48,54-58], suggesting a significant role of the band hybridization in addition to other driving forces such as magnetism and spin-orbit coupling (SOC) to account for the intriguing behaviour of TSSs in the ($MnBi_2Te_4$)$_m$($Bi_2Te_3$)$_n$ family.

# Results

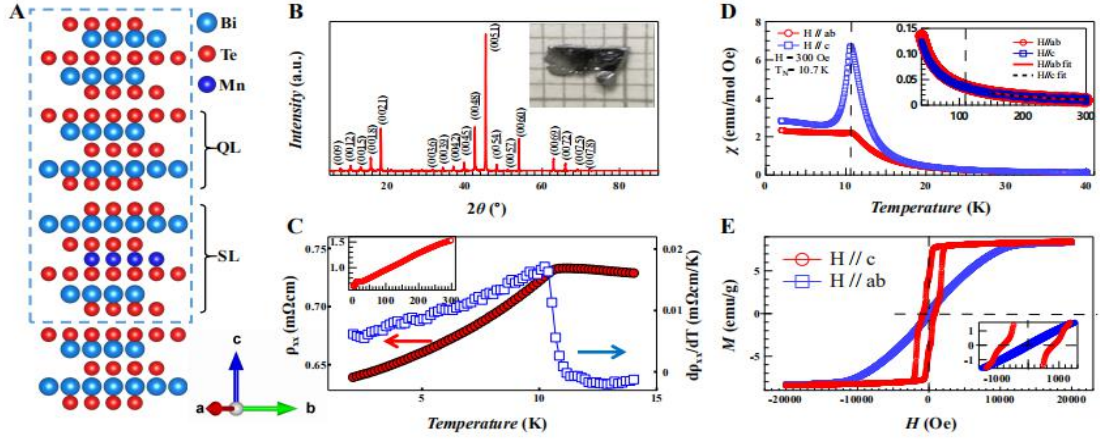

**Figure 1: Lattice structure and characterization of MnBi$_6$Te$_{10}$ single crystals.** (A), Schematic lattice structure. (B), Single crystal XRD result and peak index result. The inset shows a typical single crystal on a millimeter grid. (C), Zero-field in-plane longitudinal resistivity vs temperature. The inset shows the results up to 300 K. (D), Magnetic susceptibility vs temperature for magnetic field (300 Oe) parallel to the *ab* plane (red) and the *c* plane (blue). The inset shows the Curie−Weiss fitting for temperatures ranging from 150 K to 300 K. (E), Magnetization hysteresis at 2 K. The inset highlights the coercive fields.

MnBi$_6$Te$_{10}$ has a trigonal structure with a space group of $R\bar{3}m$. The lattice of MnBi$_6$Te$_{10}$ consists of one septuple MnBi$_2$Te$_4$ layer and two quintuple Bi$_2$Te$_3$ layers alternately stacking along the *c* axis (Fig. 1A). These SLs or QLs are coupled through weak van der Waals forces. Cleaving the single crystal perpendicular to the *c* axis could have 3 possible terminations, *i.e.,* S-termination, Q-termination and QQ-termination. The crystallinity was examined by X-ray diffraction (XRD). As shown in Fig. 1B, all peaks in the XRD pattern can be well indexed by the (00*l*) reflections of MnBi$_6$Te$_{10}$.

The zero-field in-plane longitudinal resistivity $\rho_{xx}(T)$ (Fig. 1C) shows basically a monotonic decrease with decreasing temperature for 20 K < T < 300 K, suggesting a metallic phase, which is confirmed by the following ARPES data. Around 10.7 K, the resistivity shows a weak upturn with decreasing temperature, likely from the enhanced electron scattering by magnetic fluctuation close to the AFM transition. On further cooling, the resistivity decreases again, indicating the gradual FM ordering of the spins in the *ab* plane.

The magnetic susceptibility measurement (Fig. 1D) establishes a long-range AFM order below $T_N$ = 10.7 K. A sharp cusp around $T_N$ for H//*c*, in contrast to the saturating plateau for H//*ab*, suggests that the magnetization-easy axis is the *c* direction. All these behaviours are consistent with an A-type AFM configuration (intralayer FM and interlayer AFM) along *c*-axis. Fitting the PM regime (150 - 300 K) with the Curie−Weiss formula $\chi(T) = \chi_0 + C/(T - \theta_{CW})$ gives an effective moment $\mu_{eff} \approx 5.4\mu_B/Mn$ and $5.1\mu_B/Mn$ for H//*c* and H//*ab*, respectively, confirming the high spin states of Mn$^{2+}$. The Curie−Weiss temperature is fitted as $\theta_{CW} = 9$ K and 23 K for H//*c* and H//*ab*, respectively, suggesting the existence of an additional FM interaction induced by external magnetic field. The additional FM interaction is further confirmed by the loop shape of the field-dependent magnetization [*M*(H)] results at 2 K (given in Fig. 1E). Compared to H//*c*, a much higher field is needed to fully polarize the spins with H//*ab*, indicating that the magnetization-easy axis is the *c*-axis. For the *c* direction, the spin-flip transition occurs at 0.2 T and the magnetic moments are fully polarized. This spin-flip field is much smaller than that in MnBi$_2$Te$_4$ (~3.5 Teslas) [29]. These features indicate that MnBi$_6$Te$_{10}$ is more suitable for exploring the intrinsic QAH effect.

All the above structural, magnetic and transport characterizations agree well with other works on the same materials [53,55]. Throughout this work, the ARPES spectra presented at the main text were measured at 20 K, meaning that the MnBi$_6$Te$_{10}$ samples stay at the PM phase. In addition, in Supplementary Section 1 we also present temperature-dependent ARPES measurements down to 6.5 K (AFM phase) for both S- and Q-terminations. No observable difference can be detected within the instrumental resolution [59].

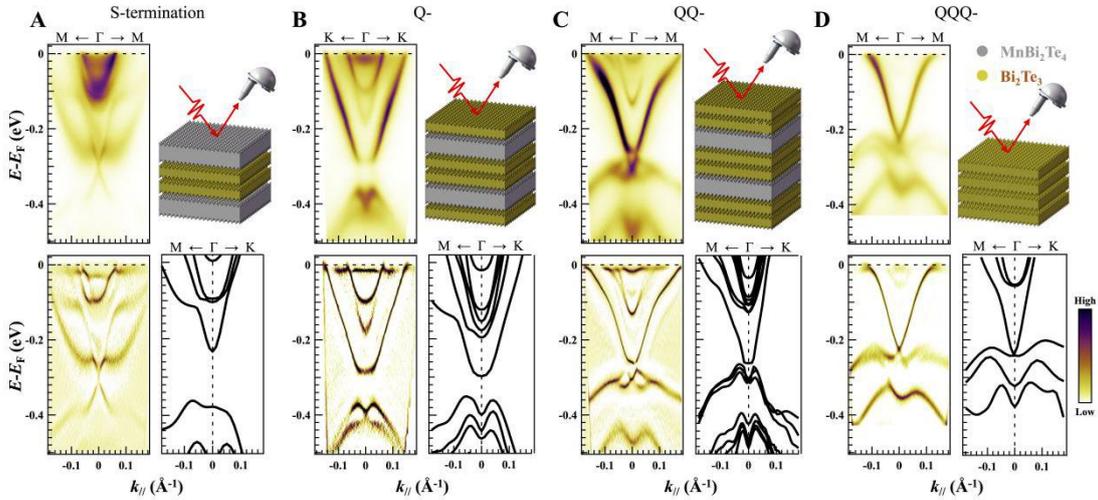

**Figure 2: Termination-dependent electronic structure of MnBi$_6$Te$_{10}$ single crystals.** (A, B, C, D) correspond to the results of S-, Q-, QQ- and QQQ-terminations, respectively. For each section, the top-left panel shows the raw ARPES spectra, while the bottom-left panel shows the 2$^{nd}$ curvature spectra. The top-right panel shows the schematic structural configuration for the corresponding DFT calculations. The bottom-right panel presents the DFT band structure, correspondingly.

To probe the intrinsic band structure of MnBi$_6$Te$_{10}$ with its different terminations, we employ a µ-Laser-ARPES system [59] with a focused laser with spot size ~5 µm to measure the cleaved MnBi$_6$Te$_{10}$ surface at 20 K. Figure 2 shows totally 4 types of distinct ARPES spectra which can be distinguished from the cleaved surfaces. Also shown are the corresponding schematics of the surface termination, second-curvature spectra and the DFT band structures. The corresponding Fermi surface mapping and the dispersion along high symmetry directions are shown in Supplementary Section 1. The assignment of the measured ARPES spectra to the corresponding terminations are based on the comparison with the DFT-calculated band structure on various slab configurations (shown in the schematics). From the ARPES spectra, each termination has characteristic band dispersion. Specifically, the S-termination shows a cordial glass-shaped conduction band minimum (CBM) and gapless linear surface states with the Dirac point located at ~300 meV below the Fermi level (Fig. 2A). The Q-termination exhibits an apparent gap of ~80 meV centered at ~320 meV below the Fermi level and an M-shaped valence band maximum (VBM) (Fig. 2B). The band structure of QQ-termination is similar to that of Q-termination but with much smaller gap (a few meVs) (Fig. 2C). Furthermore, an additional termination with generally similar band structure to QQ-termination, but gapless TSSs and less electron doping, is resolved (Fig. 2D). Its band structure can be well reproduced by a QQQ-terminated slab. This termination likely originates from either the vertical stacking disorder or the degradation of the topmost SLs. It is worthy to note that this QQQ-termination is different from $Bi_2Te_3$ single crystal in the sense that, for the latter the group velocity of TSS is ~ 3 eV·Å [60] while for the former it is only ~ 2 eV·Å.

Our PM ARPES spectra for S- and Q-terminations are quantitatively similar to the AFM ARPES spectra reported in Supplementary Section 1 and in Ref. [56-58]. This behaviour is consistent with previous works which prove that the TSSs show no observable change across the bulk Néel temperature within the experimental resolution in $MnBi_2Te_4$ [40-43]. We also calculated the total energies of several slabs with different magnetic structures (Supplementary Section 2). The total energy differences between AFM and FM states are very small (~2 meV/Mn), suggesting magnetic fluctuation being essential in affecting the macroscopic behavior [30-33]. This indicates that the underlying magnetic structure plays insignificant role in shaping the general band structure on the surfaces of this family of materials. These conclusions are in sharp contrast to the recent ARPES results reported in Ref. [20,22]. We note here that there are mainly three factors one need to caution in order to resolve the gapless TSSs using ARPES in this material family. The first one is micrometer-scale spatial resolution, to resolve termination-dependent band structure for $n = 1$ and $n = 2$ compounds; the second one is angular resolution, since the momentum window for the gapless features is as narrow as about $\pm 0.05$ Å$^{-1}$; the last but most important one is the tuning of the excitation photon energy, as sharp TSSs are only observable at certain low energy ranges duo to the matrix element effect [40].

From Figure 2, the ARPES spectra and the DFT slab band structure show general agreement, especially for the Q-, QQ- and QQQ-terminations. However, it is their detailed discrepancies such as the gapless Dirac-cone-like features at S- and Q-terminations which DFT failed to reproduce that are puzzling researchers in this field and are the main target of our efforts in this work.

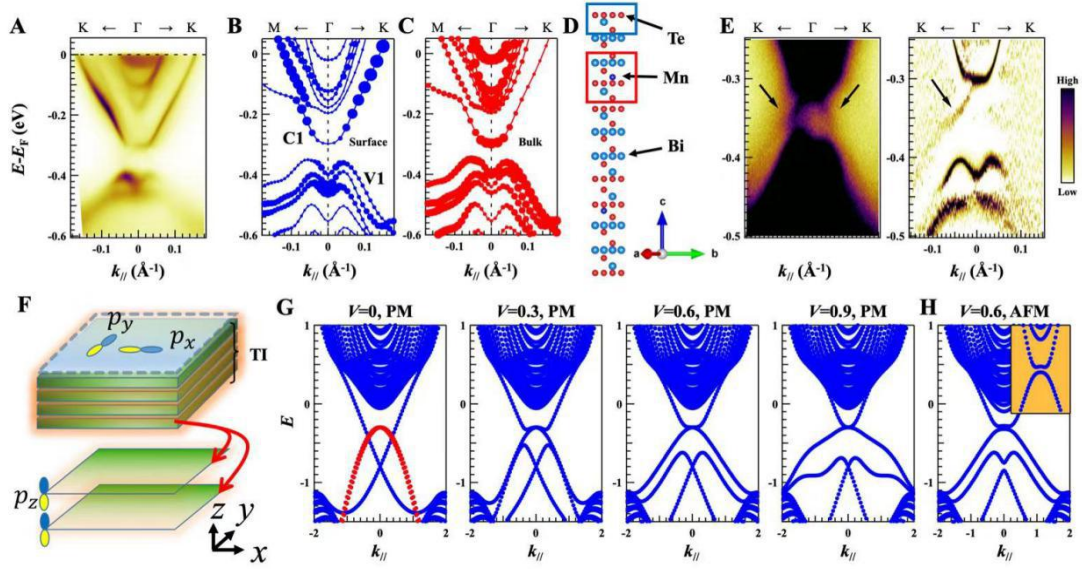

**Figure 3: Surface-bulk hybridization for the Q-termination of MnBi$_6$Te$_{10}$.** (A), the ARPES spectra of the Q-termination. (B,C,D), DFT band structure with the fatted bands projected onto corresponding layers as indicated by the blue (B) and red (C) boxes in (D). (E), Zoom in of ARPES spectra (left) and 2$^{nd}$ curvature (right) to highlight the in-gap states. (F), Schematic of the TB model. (G), TB simulated results and the band structure evolution with increasing hybridization at the PM state. The red curves represent the original RSB bands. (H), TB simulated bands with AFM order. The insets highlight the CBM splitting.

Figure 3A-E highlight the discrepancy between the experimental ARPES spectra and the DFT slab simulation for the Q-termination. Since general agreement has been reached between ARPES and DFT concerning the shape of CBM and VBM, and the gap size between them (Fig. 3A-C), experimentally the emergence of one in-gap band (indicated by black arrows in Fig. 3E), which DFT fails to reproduce, is quite unusual. The cusp of this in-gap band touches the CBM, forming a gapless Dirac-cone-like feature. Similar results have been reported by other groups at the Q-terminations in MnBi$_4$Te$_7$ [54,56] and MnBi$_6$Te$_{10}$ [56,58].

We firstly try to track down the origin of the apparent gap. As shown by the DFT calculated bands and the corresponding projection to different layers in Fig. 3B-D, the lowest conduction band (C1) is mainly surface-originated, with surface domination increasing with the wave vector. At the Γ point, however, C1 is dominated by bulk states. For the highest valence band (V1), the opposite occurs. The surface states dominate around Γ and bulk contribution increases with wave vector. This demonstrates a surface-bulk band hybridization and gap opening due to the "avoided crossing" [61]. We then proceed to use a tight binding (TB) model analysis to show that this hybridization can indeed lead to the emergence of the apparent gap and the additional in-gap states (Fig. 3E).

For simplicity, we consider a scenario where the TSS Dirac cone hybridize with additional quasi-two-dimensional Rashba-split bulk bands (RSB). These RSBs are not intrinsically related to the underlying topological insulating phase. They originate from bulk bands which gain surface characteristic due to their proximity to the surface, similar to the quantum well states observed at the surfaces of typical TI $Bi_2Se_3$ family [44]. Traces of this identification can indeed be found in the DFT calculations (Fig. 3B and C). As schematically drawn in Fig. 3F, the TI is modeled by a 3D stacking of bilayers of $p_z$-like Wannier orbitals, which has been shown to be appropriate for this series of TIs [24]. The RSBs may originate from either certain additional $p_z$ orbitals, or certain $p_x/p_y$-like Wannier orbitals which lie close in energy. In our simplified model, we consider spin-orbit coupled $(p_x + ip_y)\uparrow$ and $(p_x - ip_y)\downarrow$ Wannier orbitals in a fictitious layer. Note that they are the only *p*-wave states that could be distinguished from the TI $p_z$-orbitals under the $C_3$ symmetry at the surface of the TI. Nevertheless, the results are qualitative similar if the RSBs come from $p_z$-like orbitals instead, as we shall mention later. The RSB orbitals couple with the topmost bilayer of the TI, and the effective Hamiltonian reads,

$$H = H_{TI} + H_{RSB} + H_{hyb}.$$

In order, the three terms stand for the Hamiltonian of the TI, the RSBs, and the hybridization. While more details are provided in the Supplementary Section 3, the hybridization term deserves a special attention. Expanded around the Γ-point, it reads,

$$H_{hyb} = V \sum_{m,k} (-1)^m \left[ (k_x - ik_y) c^\dagger_{0mk\uparrow} d_{k\uparrow} + (k_x + ik_y) c^\dagger_{0mk\downarrow} d_{k\downarrow} \right] + H.c.,$$

where, $V$ denotes the strength of the hybridization, $c^\dagger_{0m}$ and $c_{0m}$ represent the creation and annihilation of $p_z$-orbitals on the top TI bilayer where $m = 0,1$ index its two sublayers, and, $d^\dagger_{\uparrow(\downarrow)}$ and $d_{\uparrow(\downarrow)}$ create and annihilate the respective $(p_x + ip_y)\uparrow$ and $(p_x - ip_y)\downarrow$ orbitals of the RSBs. It is important to note that the time-reversal and $C_3$ symmetry ensures that the hybridization vanishes at the Γ point, meaning that the Dirac crossings of the RSBs and that of the TSSs remain intact. Due to the finite hybridization away from the Γ point, the two Dirac crossings switch designations, *i.e.*, the original RSB crossing become the crossing of the new Dirac cone that connects the conduction and valence bands, and vice versa. In this manner, the essential topological properties of the hybrid system are unaltered from the original TI. As a side remark, had we considered RSBs originating from other orbitals, such as one with $p_z$-symmetry, the hybridization between the two Dirac crossings at the Γ point becomes finite, shifting their relative energy. However, this hybridization does not lift the degeneracy of the individual crossings so long as time reversal symmetry is preserved. Hence, as in the above case, the two crossings switch designations and the overall topology remains.

In Fig. 3G, we present the results of typical TB simulations of the Q-termination within the PM state, in which only the hybridization strength $V$ is varied. Note that in this case a small Rashba SOC is taken for the RSBs in order to obtain a good fit. For $V = 0$, the TSSs (blue) and RSBs (red) remain intact. With increasing $V$, the main features of the experimental spectra are readily reproduced, including the M-shaped VBM, the hybridization gap and the hole-type in-gap states.

Further inclusion of AFM order (Fig. 3H) can even reproduce the small gap in the original Dirac cone observed by ARPES here (Fig. 3E), and the tiny splitting at the CBM, which is observed by others [58].

Within this picture, those intriguing features found in the experimental spectra of Q-terminations (Fig. 3E) can now be understood. The M-shaped VBM consists of the original TSS Dirac cone and RSBs; the apparent gap opens in the upper Dirac cone region due to the surface-bulk hybridization; the gapless feature comes from the symmetry-protected touching of the gapped TSSs and split RSBs at the Γ point. With AFM order, gaps are opened at both original TSS Dirac cone and the new touching point, but the gap size could be so small that extrinsic factors can smear it easily.

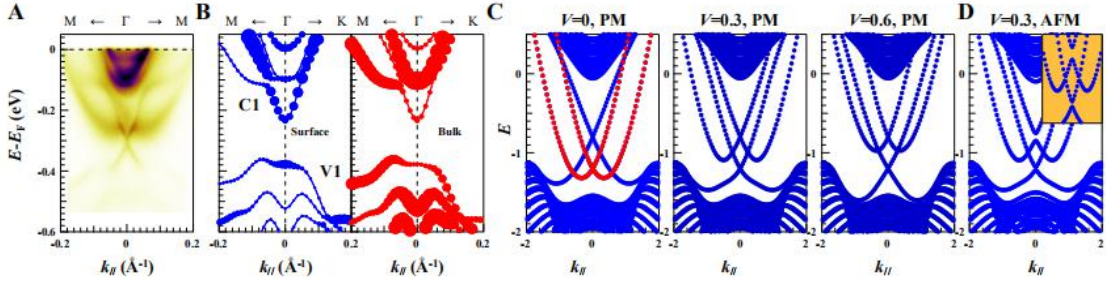

**Figure 4: Surface-bulk hybridization for the S-termination of $MnBi_6Te_{10}$.** (A), the ARPES spectra. (B), DFT band structure with the fatted bands projected onto corresponding surface (blue filled circles) and bulk layers (red filled circles). (C), TB model simulated band structure and its evolution with increasing hybridization at PM state. The red curves represent the original RSB bands. (D), TB simulated bands at AFM order. The inset highlights the tiny gap of the new TSSs.

This TB hybridization model is also applicable to the band structure of S-termination shown in Fig. 4. Similarly, from DFT, the C1 band is surface-dominated and V1 is bulk-dominated while around the Γ point the opposite happens, suggesting a surface-bulk hybridization picture (Fig. 4B). In Fig. 4C, we use the TB model presented above to model the S-termination bands. This time CBM-like RSBs hybridize with the lower Dirac cone of TSSs. With appropriate SOC and increasing $V$, the cordial glass-shaped CBM and gapless linear Dirac cones start to manifest. The $V = 0.3$ panel matches the experimental spectra the most. In Fig. 4D, the AFM order is included in the simulation and tiny gaps opens in the new Dirac cone, in agreement with ARPES results [58]. The consequent (nearly) gapless features at S-termination thus come from the original RSB Dirac cone. The original TSS Dirac cone gains broadness due to its hybridization with bulk states and its tiny gap at AFM state is smeared.

We have demonstrated the applicability of a surface-bulk hybridization model to account for the band structure at the S- and Q-terminations of $MnBi_6Te_{10}$. Built on the DFT orbital projection analyses, this model is well-grounded and provides a unique physical picture to explain the (nearly) gapless behaviour of TSSs for all terminations. In this picture, the Dirac cones identified by ARPES are of bulk origin, with wave functions extending much more deeply into the crystal than the non-hybridized ones. This means that the impact of $S$ symmetry breaking at the surface on these spectral features is dramatically reduced, naturally explaining their (nearly) gapless behaviour at all surfaces within AFM phase. Due to their similarity in the lattice structure and band structure, the universal (nearly) gapless behaviour from all terminations of $(MnBi_2Te_4)_m(Bi_2Te_3)_n$ can be understood now in this picture.

## MATERIALS AND METHODS

### Sample growth and characterization

High quality $MnBi_6Te_{10}$ single crystals were grown by the conventional high-temperature solution method with $Bi_2Te_3$ as the flux. Mn (purity 99.98%), Bi (purity 99.999%) and Te (99.999%) blocks were placed in an alumina crucible with a molar ratio of Mn: Bi: Te = 1: 11.3: 18. Then the alumina crucible was sealed in a quartz tube under the argon environment. The assembly was first heated up in a box furnace to 950 ℃, held for 10 hrs, then subsequently cooled down to 700 ℃ over 10 hrs and further cooled down slowly to 575 ℃ over 100 hrs. After this heating procedure, the quartz tube was taken out quickly and then decanted into the centrifuge to remove the excess flux from the single crystals.

Resistivity measurement were performed by a Quantum Design (QD) Physical Properties Measurement System (PPMS) with a standard six-probe method. The driven current is 10 mA and flows in the *ab* plane. Magnetic measurements were performed using the QD PPMS with the Vibrating Sample Mangetometer (VSM) mode. Temperature dependent magnetization results were collected with an external magnetic field of 300 Oe, both along and perpendicular to the *c*-axis direction of the sample.

### ARPES measurement

μ-Laser-ARPES measurements were performed at the Hiroshima Synchrotron Radiation Center (HSRC), Hiroshima, Japan with a VG Scienta R4000 electron analyzer and a photon energy of 6.3 eV [59]. The energy and angular resolution were better than 3 meV and less than 0.05°, respectively. Samples were cleaved *in situ* along the (001) crystal plane under ultra-high vacuum conditions with pressure better than $5 \times 10^{-11}$ mbar and temperatures below 25 K.

### DFT Calculation method

First-principles calculations were performed using the Vienna *ab initio* simulation package (VASP) [62,63] within the framework of density-functional theory [64,65]. The generalized gradient approximation (GGA) with the Perdew-Burke-Ernzerhof (PBE) formalism was chosen for the exchange-correlation functional [66]. The projector augmented wave (PAW) method was implemented to treat core-valence interactions with a cutoff energy of 400 eV for the plane-wave expansion [67,68]. The full Brillouin zone was sampled by a $8 \times 8 \times 1$ Monkhorst-Pack grid [69]. The slab structures (including both lattice and fractional coordinates) were fully relaxed until the forces on each atom were less than 0.01 eV/Å. Considering the strongly correlated nature of 3*d* electrons in Mn, we introduced on-site Coulomb repulsion by employing GGA+U calculations [70] and set the correlation energy to be 5 eV, which worked well in the previous works [71].

**Tight binding simulation method**

We construct a model to effectively simulate the hybridization between the TI surface states and a pair of Rashba-split bulk bands. Since we are only concerned about the physics around the Γ-point, it suffices to build the model on a lattice with tetragonal symmetry to keep the expressions in their simplest forms. No essential feature of the hybridization in question is lost due to this simplification. We generalize a standard *k·p* theory of TI [72] to a 3D lattice model, which is formed by a stacking of bilayers with each sublayer featuring a $p_z$-orbital. Regarding the hybridization to bulk states, we consider that the topmost bilayer of the TI is coupled to an extra band with Rashba SOC, whose band top (or bottom) overlaps with the TSSs in momentum space. More details can be found in the Supplementary Section 3.

**ACKNOWLEDGEMENTS**

We thank Qihang Liu, Junhao Lin, Liusuo Wu, Weiqiang Chen and Haizhou Lu for helpful discussions. J.F.S is supported by the National Foundation for Young Scientists of China (Grant No. 11804402). C.L. is supported by the National Natural Science Foundation of China (NSFC) (No. 11504159 and No. 11674149), NSFC Guangdong (No. 2016A030313650), the Guangdong Innovative and Entrepreneurial Research Team Program (No. 2016ZT06D348), the Shenzhen Key Laboratory (Grant No. ZDSYS20170303165926217), and the Technology and Innovation Commission of Shenzhen Municipality (Grants No. JCYJ20150630145302240 and No. KYTDPT 20181011104202253). X.J.Z is supported by the National Key Research and Development Program of China (Grant No. 2016YFA0300300 and 2017YFA0302900) and the National Natural Science Foundation of China (Grant No. 11888101). H.X. is supported by Center for Computational Science and Engineering at Southern University of Science and Technology. B.S. is supported by the Hundreds of Talents program of Sun Yat-Sen University, the Fundamental Research Funds for the Central Universities. W.H. thank the startup grant at SUSTech. The ARPES measurements were performed with the approval of the Proposal Assessing Committee of the Hiroshima Synchrotron Radiation Center (Proposal Numbers: 19AU002, 19AU007 & 19BG006).